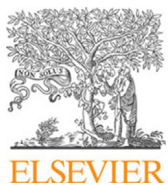
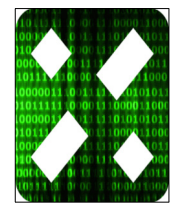

Original software publication

# Evoplex: A platform for agent-based modeling on networks

Marcos Cardinot [a,*], Colm O'Riordan [a], Josephine Griffith [a], Matjaž Perc [b,c,d]

[a] *Discipline of Information Technology, National University of Ireland, Galway, Ireland*
[b] *Faculty of Natural Sciences and Mathematics, University of Maribor, Koroška cesta 160, SI-000 Maribor, Slovenia*
[c] *CAMTP – Center for Applied Mathematics and Theoretical Physics, University of Maribor, Mladinska 3, SI-2000 Maribor, Slovenia*
[d] *Complexity Science Hub Vienna, Josefstädterstraße 39, A-1080 Vienna, Austria*

**ABSTRACT**

Agent-based modeling and network science have been used extensively to advance our understanding of emergent collective behavior in systems that are composed of a large number of simple interacting individuals or agents. With the increasing availability of high computational power in affordable personal computers, dedicated efforts to develop multi-threaded, scalable and easy-to-use software for agent-based simulations are needed more than ever. Evoplex meets this need by providing a fast, robust and extensible platform for developing agent-based models and multi-agent systems on networks. Each agent is represented as a node and interacts with its neighbors, as defined by the network structure. Evoplex is ideal for modeling complex systems, for example in evolutionary game theory and computational social science. In Evoplex, the models are not coupled to the execution parameters or the visualization tools, and there is a user-friendly graphical interface which makes it easy for all users, ranging from newcomers to experienced, to create, analyze, replicate and reproduce the experiments.

## Code metadata

| | |
|---|---|
| Current code version | v0.2.1 |
| Permanent link to code/repository used for this code version | https://github.com/ElsevierSoftwareX/SOFTX_2018_211 |
| Legal Code License | Apache 2.0 (EvoplexCore) and GPLv3 (EvoplexGUI) |
| Code versioning system used | Git |
| Software code languages, tools, and services used | C++, OpenGL, Qt and CMake. |
| Compilation requirements, operating environments & dependencies | C++ compiler (e.g., GCC, Clang or MSVC), Qt Framework and CMake. |
| If available Link to developer documentation/manual | https://evoplex.org/api |
| Support email for questions | evoplex@googlegroups.com |

## Software metadata

| | |
|---|---|
| Current software version | v0.2.1 |
| Permanent link to executables of this version | https://github.com/evoplex/evoplex/releases |
| Legal Software License | GPLv3 |
| Computing platforms/Operating Systems | Linux, OS X, Microsoft Windows and Unix-like |
| Installation requirements | OpenGL 2.0+ |
| If available, link to user manual - if formally published include a reference to the publication in the reference list | https://evoplex.org/docs |
| Support email for questions | evoplex@googlegroups.com |

## 1. Motivation and significance

Agent-based modeling (ABM) has been used as a framework to simulate complex adaptive systems (CAS) in a wide range of domains such as life sciences, ecology and social sciences [1–7]. Those systems are composed of a number of interacting agents each of whom have a defined set of attributes and can exhibit specific behaviors based on their interactions with the environment and the other agents [8]. Research in this field usually aims to explore how small changes in individual behavior can both affect and promote collective behavior throughout the system [9–11].

* Corresponding author.
*E-mail address:* marcos.cardinot@nuigalway.ie (M. Cardinot).

Given the flexibility of the ABM approach and the increasing computing power of cheap personal computers, efforts to develop reusable, flexible, multi-threaded, scalable and user-friendly software are more than ever required by the scientific community. However, despite the high number of existing ABM toolkits and platforms available [12], due to the heterogeneity and diversity of the areas of research and application domains, most researchers still prefer to implement individual and domain-specific, specialized software from scratch, which is usually not publicly released. Many researchers write MATLAB or *Mathematica* based scripts which, although being complete and well-known scientific platforms, are neither free nor open-source, which therefore reduces the transparency and re-usability of the developed models [13].

In fact, implementing a highly specialized solution from scratch is time-consuming, complex and error-prone. Many projects try to overcome this by implementing a toolkit or platform for a general purpose problem domain. For instance, some projects such as NetLogo [14] and GAMA [15] succeed in providing generic and reusable software; however, they require the user to learn their specific programming language. A wide range of the ABM solutions including MASON [16] and Repast [17] are written in Java [18], which make them cross-platform and usually faster than some Python or JavaScript alternatives like Mesa [19] and AgentBase [20]. However, they usually require modelers to be highly proficient in the language or they have critical scalability issues. Overall, the main issues with some existing software include the use of old/deprecated technologies, not following state of the art in software engineering, developing single-threaded applications and not being community-friendly.

Furthermore, despite being a common strategy in the field, many ABM projects start with the promising and challenging intention of developing powerful software to meet any requirement in the field, from simple cellular automaton models to complex and realistic geographical information science (GIS) models. Unfortunately, this promising approach usually results in making the code base very complex and hard to both optimize and maintain. In reality, given the small size of the development teams, there is no best strategy for all scenarios, and the user choice is usually guided by their familiarity with the languages or technologies used in the software. In this way, defining a clear and focused scope can help solve those issues.

Thus, in this paper we present Evoplex, a cross-platform, free and open-source software which aims to mitigate the issues outlined by providing a fast and fully modular ABM platform for implementing and analyzing models which impose an explicit graph-theoretical approach, i.e., the agents are represented as nodes (or vertices) and edges (or links) represent connections between agents.

## 2. Software description

Evoplex is a fast, robust and extensible platform for developing agent-based models and multi-agent systems on networks. Here, each agent is represented as a node in the network and is constrained to interact only with its neighbors, which are linked by edges in the network. Although not limited to a particular domain, Evoplex is ideal for tackling problems associated with evolutionary computation, complex adaptive systems, evolutionary game theory and cellular automata.

As shown in Fig. 1, the Evoplex workflow is very straightforward and intuitive. The engine processes *projects* as inputs. A **project** is a plain table (csv file) where the experiments are listed along the rows, and the inputs to each experiment are placed along the columns. An **experiment** is defined by a set of parameter settings (inputs) necessary to perform one **trial** (simulation) and (optionally) the required data outputs, which can be the result of some statistical function and/or the state of the set of nodes/edges for each time step. Each experiment can run for one or more trials, i.e., repeat the same experiment using different pseudo-random generator seeds. The strategy of having the projects defined in plain text files aims to make it easier for users to replicate and reproduce their results. Furthermore, it allows newcomers to interact with the models without requiring any programming skills.

We provide a user-friendly and interactive graphical user interface (GUI) to allow for creating, opening, running and saving projects. Also, the GUI provides many useful tools including interactive graph and grid views. Moreover, Evoplex allows several experiments to run at the same time. These are automatically distributed in parallel across multiple threads.

### 2.1. Software architecture

Evoplex is simple, user-friendly and was built with performance in mind from the start. It is cross-platform and runs on every major platform, i.e., Linux, Microsoft Windows, and MacOS. Evoplex is developed in modern C++ 14, based on Qt, which is one of the most popular and successful C++ frameworks available to date. Moreover, Evoplex includes CMake scripts to ease the compilation and setup from its source code.

The Evoplex application bundles three main open-source components: the *EvoplexCore* library, the *EvoplexGUI* library and a collection of plugins (example models and graph generators). The *EvoplexCore* library is available under the Apache 2.0 License, which is permissive, free and commercially friendly. The *EvoplexGUI* library is available under the GNU GPLv3 license, which is also free but is conditioned on making the source code of licensed works and modifications available.

Following a common practice in software engineering, the Evoplex architecture is guided by a fully modular approach. The core component, *EvoplexCore*, splits its implementation into both private and public Application Programming Interfaces (APIs). The private API is intended for internal use only and is where the simulations will actually occur; it is responsible for managing the I/O operation, parsing inputs, handling the CPU threads and memory, loading and creating instances of plugins and others. The public API exposes all the tools and services needed to develop a plugin, which can be either a model or a graph generator.

Fig. 2 shows a simplified overview of the overall software architecture, which is composed of four major layers: the kernel (i.e., *EvoplexCore* library), the plugins, the data and the applications layers. The current version of the Evoplex application layer includes *EvoplexGUI*, which implements a graphical user interface on top of *EvoplexCore* (kernel) to provide a number of interactive and user-friendly tools. Note that as the kernel is completely independent of the applications layer, Evoplex can be distributed with different user-interfaces but share the same engine (kernel). For instance, one may want to implement an *EvoplexCLI* application to perform simulations via command-line, or an *EvoplexWeb* application to provide visualization tools on a web browser.

In Evoplex, every model or graph is a plugin and is compiled independently of the main application. The creation of plugins is very straightforward and requires a very basic knowledge of C++. Given the Evoplex approach of not coupling the visualization tools nor inputs/outputs to the model, the models' code is usually very simple and short. We provide a few examples of plugins of easy reuse and customization.[1] In summary, a plugin comprises four files: *CMakeLists.txt* which does not need to be changed by the modeler and is just a CMake script to ease the compilation process and make it portable across different compilers and IDEs, the *plugin.cpp* (source) and *plugin.h* (header) files where the modeler implements the model's algorithm, and *metadata.json* which holds the definition of all the attributes of the model.

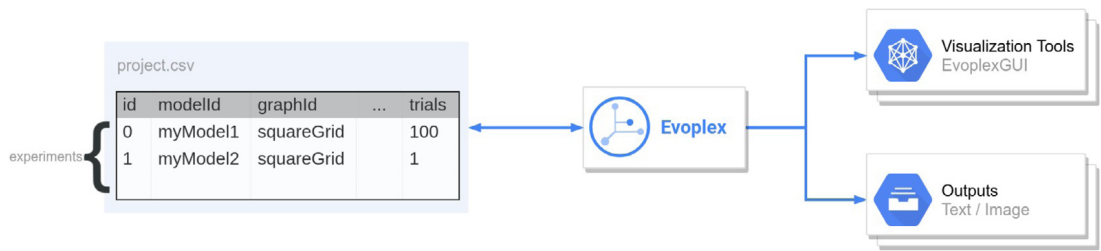

**Fig. 1.** Simplified overview of the user workflow.

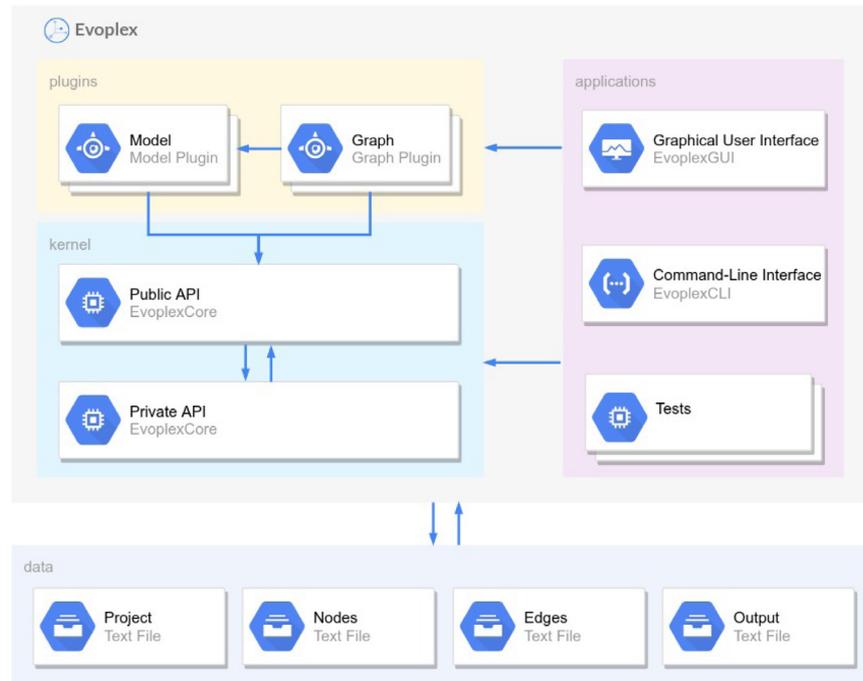

**Fig. 2.** Simplified illustration of the software architecture.

Moreover, Evoplex uses automated Continuous Integration (CI) to make sure that the code base works as expected and to allow early detection of problems. After every new commit, the CI system automatically builds Evoplex from the source code, executes regression tests, unit-tests and static code analysis across a range of different versions/distributions of the supported platforms, i.e., Linux, Microsoft Windows, and MacOS.

### 2.2. Software functionalities

The Evoplex application comes with a user-friendly and intuitive GUI that allows loading and unloading of plugins at runtime and provides a bunch of widgets and tools to allow for the creation and running of experiments and for analyzing (or visualizing) their outputs. The main tools and widgets are described below:

---

[1] https://evoplex.org/docs/example-plugins.

- *Project View*: As shown in Fig. 3, when opening a project, all experiments are listed in a table which is dynamic and customizable and allows running, pausing and queuing multiple experiments at the same time. When running the experiments, Evoplex automatically manages the available resources to run them as fast as possible (in parallel) and use as little memory as possible.
- *Experiment Designer*: This widget is displayed beside the *Project View* in Fig. 3 and allows creating, removing and editing of experiments.
- *Nodes Generator*: This tool can be accessed in the *Experiment Designer* and provides handy functions to ease the generation of the initial set of nodes.
- *Experiment View*: This widget is opened when the user double-clicks on an experiment in the *Project View*. It allows for the opening of multiple visualization tools at the same time, which can be set to show different trials of the same experiment. For instance, given an experiment with a stochastic model which runs for 10 trials; the user may want to visualize the behavior of the trials side by side to investigate the effects of randomness over time.
- *Graph/Grid View*: Evoplex provides both graph (nodes and edges — Fig. 4) and grid (cells) views. Those views allow zooming in and out, exporting nodes/edges as a text file, taking screenshots, selecting a node to inspect and change the state of its attributes and others. Also, it allows changing the nodes/edges size and choosing which attribute and colormap will be represented in the nodes/edges.

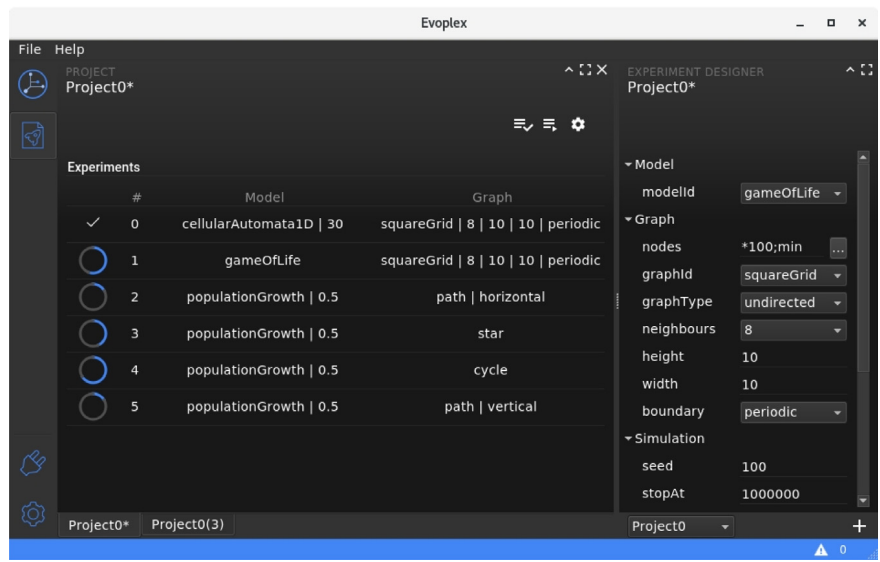

**Fig. 3.** Screenshot of Evoplex 0.2.1 showing the *Project View* and the *Experiment Designer* tools.

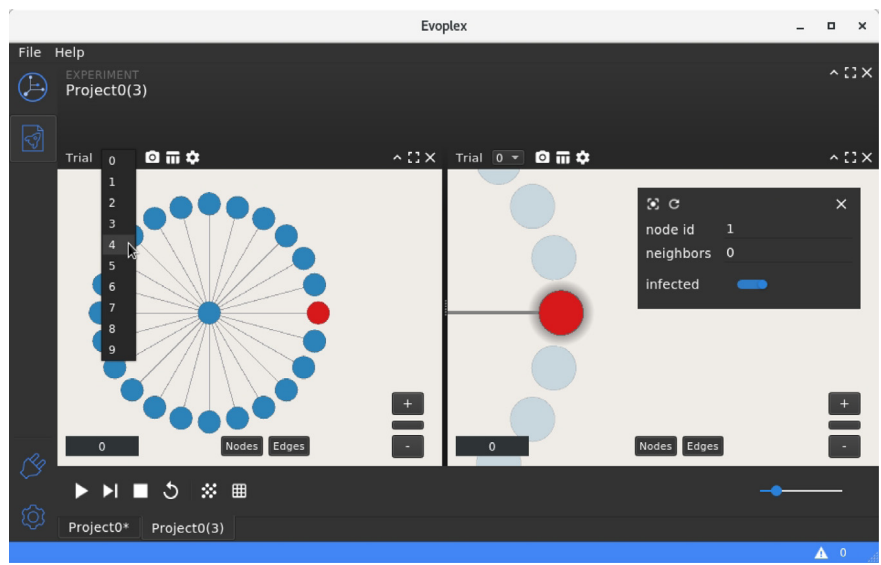

**Fig. 4.** Screenshot of Evoplex 0.2.1 showing the *Experiment View* docking two instances of the *Graph View* at different positions.

Differing from most of the other ABM solutions (e.g., NetLogo [14], MASON [16] and GAMA [15]), in Evoplex, the widgets are not statically coupled to the model plugin. That is, the model plugin only defines the entities' attributes and implements the algorithm to describe the nodes' (agents) behavior for each time step. Then, at runtime and not requiring any programming skill, the users have the freedom to decide which widgets they want to open and where they want to place them. Also, all widgets can be detached from the main window, enabling users to open different views in multiple monitors or attach them at different positions and sizes in the screen.

### 3. Illustrative examples

In order to illustrate the use of Evoplex, we consider an implementation of the widely known model of a spatial prisoner's dilemma (PD) game proposed by Nowak & May in 1992 [21]. In the PD game, agents can be either cooperators or defectors, and receive a fixed payoff based on a pairwise interaction. That is, given two agents, if both are cooperators, both get a reward $R = 1$; if both are defectors, both get a punishment $P = 0$; if a cooperator plays with a defector, the cooperator receives $S = 0$, and the defector receives $T$ (temptation to defect) [22].

In this model, agents (nodes) are placed in a square grid, where, in each round: every node accumulates the payoff obtained by playing the PD game with all its immediate neighbors and itself; then, each agent copies the strategy of the best performing agent in its neighborhood, including itself. Note that the model's source code is also freely available online[2] under the MIT License terms.

Fig. 5 shows a screenshot of an experiment created with the *Experiment Designer* tool, using an implementation of the PD model in Evoplex. To reproduce this output, run the experiment for one step, open the *Grid View* and place a single defector

---

[2] https://github.com/evoplex/model-prisonersDilemma.

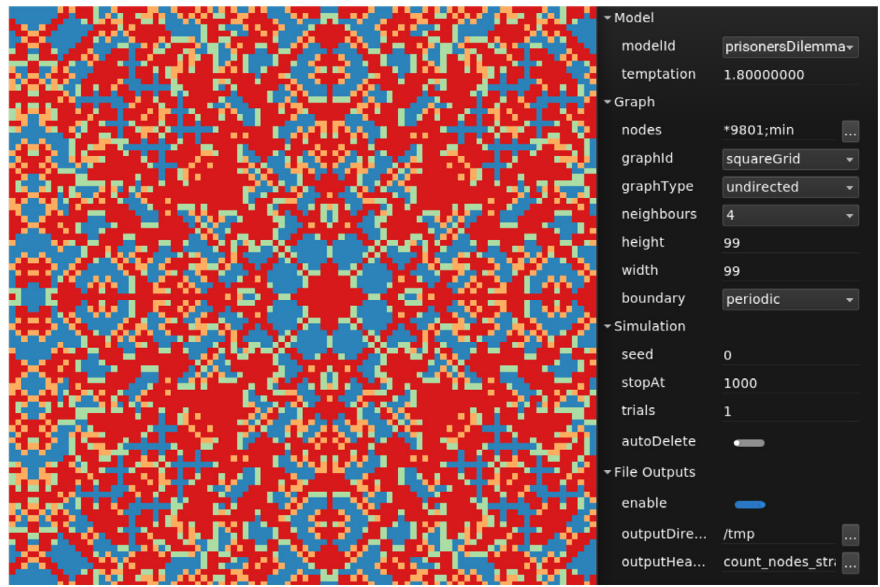

**Fig. 5.** In this experiment, the model (*prisonersDilemma*) is set with a *temptation to defect* equal to 1.8; the graph is initialized with a single defector (*strategy* = 1) at the center of a 99 × 99 *squareGrid* with periodic boundary conditions, fully populated with cooperators (*strategy* = 0), undirected edges and von Neumann neighborhood; the simulation is fed with a pseudo-random generator seed equal to 0 (which does not make any difference in this fully deterministic model), and is set to run for a maximum of 1000 time steps for only one trial; finally, it also stores the frequency of each type of strategy over time. In the *Grid View*, the colors blue, red, green and yellow corresponds to cooperators, defectors, new cooperators and new defectors respectively (for interpretation of the references to color in this figure legend, the reader is referred to the web version of this article).

(*strategy* = 1) in the middle of the grid. Then, when running the experiment for more steps, it is possible to observe the emergence of chaotically changing spatial patterns as reported by Nowak & May [21].

## 4. Impact

Evoplex is intended to address research whose methodology comprises a simulation-based approach to evolve outcomes of populations of autonomous and interacting agents. It has been used to support research in a number of areas, including spatial game theory and evolutionary game theory [1,23,24]. In those scenarios, agents are described in terms of graph theory, i.e., a graph (network) consisting of a set of nodes (agents) and edges (agents' connections).

Despite having a few options of agent-based modeling (ABM) software available, none of them are really suitable for this area of research. Beyond the issues mentioned in Section 1, most of the existing simulators have very limited performance and are unable to handle the complexity of the models which are investigated at present, e.g., coevolutionary models with a large number of agents. Thus, one of the main impacts and contributions of Evoplex to this field of research is to provide an easy-to-use and high-performance platform for simulating large-scale experiments.

Moreover, another recurring issue with existing ABM software is that they are designed to run and analyze one experiment at a time. However, research in the field usually needs to explore the outcomes of large populations for a wide range of parameter settings, which in many cases require many Monte Carlo steps to converge. In this case, the user needs to modify the model's source code or write a script on top of it to automate the execution of the experiments, which will usually run in a single thread, one at a time. Some projects like FLAME [25] and OpenMOLE [26] succeeded in allowing efficient parallel processing, but their use and configuration are not straightforward. Thus, in those cases, we observed that for any small interaction with the model, the user ends up having to change the code/script back and forth very often, which is both error prone and difficult for non-experienced users.

Evoplex changes the paradigm of ABM for graphs by allowing nodes and edges to be independent entities. Thus, given a set of nodes (agents), the user can easily investigate how changes in the topology may affect the population's behavior (and vice versa) without touching the source code or changing the model. Also, the robust and multi-threaded engine combined with the user-friendly GUI makes it easier for users to share, reproduce, replicate and analyze the experiments. Evoplex is free, non-profit and is fully open-source with a permissive license, allowing for both commercial and academic use.

## 5. Conclusions

We have presented Evoplex, a flexible, fast and multi-threaded platform for agent-based modeling imposing an explicit graph-theoretical approach. We discussed that, different to other software, in Evoplex, the models are not coupled to the execution parameters nor the visualization tools. Also, it provides a user-friendly GUI which makes it easy for all users, ranging from newcomers to experienced, to create, analyze, replicate and reproduce experiments. As an open-source project, we encourage users to provide feedback, share models and contribute to improving the software. Evoplex is an ever-evolving project, and future work will involve adding support for multilayer networks, as well as implementing more plugins, and developing more visualization widgets for the GUI.

### Acknowledgments

This work was supported by the National Council for Scientific and Technological Development (CNPq-Brazil) (Grant 234913/2014-2), and by the Slovenian Research Agency, Slovenia (Grants J4-9302, J1-9112 and P1-0403).

### Competing interests

The authors declare no competing interests.